\documentclass[twocolumn,aps,floats,letterpaper,floatfix,groupedaddress]{revtex4-1}
\usepackage{amsmath}
\usepackage{graphicx}
\usepackage{amsfonts}
\usepackage{amssymb}
\usepackage{bm}
\usepackage{epsfig,float,afterpage}
\usepackage[colorlinks,linkcolor=blue,citecolor=blue,urlcolor=blue]{hyperref}
\usepackage[usenames,dvipsnames]{color}

\newcommand{\beq}{\begin{equation}}
\newcommand{\eeq}{\end{equation}}
\newcommand{\bes}{\begin{subequations}}
\newcommand{\ees}{\end{subequations}}
\newcommand{\bea}{\begin{eqnarray}}
\newcommand{\eea}{\end{eqnarray}}
\newcommand{\ba}{\begin{array}}
\newcommand{\ea}{\end{array}}
\newcommand{\beqn}{\begin{eqnarray*}}
\newcommand{\eeqn}{\end{eqnarray*}}

\newcommand{\f}[2]{\frac{#1}{#2}}

\newcommand{\om}{\omega}

\newcommand{\la}{\langle}
\newcommand{\ra}{\rangle}

\newcommand{\dg}{\dagger}

\def\nn{\nonumber}

\begin{document}
\title{Critical features of nonlinear optical isolators for improved nonreciprocity}
\author{Dibyendu Roy}
\affiliation{Raman Research Institute, Bangalore 560080, India}
\begin{abstract}
  Light propagation in a nonlinear optical medium is nonreciprocal for spatially asymmetric linear permittivity. We here examine physical mechanism and properties of such nonreciprocity (NR). For this, we calculate transmission of light through two models of a nonlinear optical isolator consisting of (a) a two-level atom and (b) a driven $\Lambda$-type three-level atom coupled asymmetrically to light inside open waveguides. We find a higher NR in the model (b) than in the model (a) due to a stronger optical nonlinearity in the former. We determine the critical intensity of incident light for maximum NR and a dependence of the corresponding NR on asymmetry in the coupling. Surprisingly, we find that it is mainly coherent elastic scattering compared to incoherent scattering of incident light which causes maximum NR near the critical intensity. We also show a higher NR of an incident light in the presence of an additional weak light at the opposite port.
\end{abstract}

\maketitle
\section{Introduction}
Light propagation is nonreciprocal when transmission of light is different under reversal of incoming light's direction. Nonreciprocity (NR) in light propagation can be achieved using various physical mechanisms including magneto-optical Faraday rotation \cite{Gauthier86, BiNat2011}, parametric modulations \cite{Yu2009, KamalNat2011, SounasNatC2013, EstepNat2014}, optical nonlinearity plus spatially asymmetric linear permittivity \cite{RoyPRB2010, RoyNatS2013, FratiniPRL2014, Yu15, Shi2015, FratiniPRA2016} and spin-orbit interaction of light \cite{Mitsch14, Petersen14, Le15}. Optical NR without magnetic materials and fields has attracted a lot of interest in the recent years for its  suitability in an on-chip integration of an optical isolator \cite{Yu2009, KamalNat2011, SounasNatC2013, EstepNat2014, RoyPRB2010, RoyNatS2013, FratiniPRL2014, Yu15, Shi2015, FratiniPRA2016, Mitsch14, Petersen14, Le15, Miroshnichenko10, Fan12, Bender2013, RodrguezScience2013, Peng2014, Chang2014, SayrinPRX2015}.

A few years ago, we proposed an all-optical diode or isolator \cite{RoyPRB2010} in a simple system consisting of a two-level atom (2LA) being asymmetrically coupled to light inside one-dimensional (1D) waveguides, such as superconducting transmission lines \cite{Astafiev10a,Hoi11} and line-defects in photonic crystals \cite{Lodahl13}. A propagating light inside such open waveguides can be tightly confined to deeply subwavelength sizes in the transverse dimensions. It leads to an effective photon-photon correlation (optical nonlinearity) through strong atom-photon coupling even at a lower light power \cite{Roy2017}. The NR in the transmission of light in the proposed diode is achieved via (i) optical nonlinearity which results in an incoming light's power-dependent dielectric response of the system and (ii) asymmetric coupling which creates a spatially asymmetric linear permittivity across the atom. Asymmetric permittivity causes a spatially asymmetric dielectric response. We have shown that while single-photon transmission is the same under reversal of incoming light's direction, the two-photon transmission is not \cite{RoyPRB2010}. This mechanism has been investigated in many recent studies \cite{RoyNatS2013, FratiniPRL2014, Yu15, FratiniPRA2016} for nonreciprocal transmission. 

\begin{figure}
\includegraphics[width=0.90\linewidth]{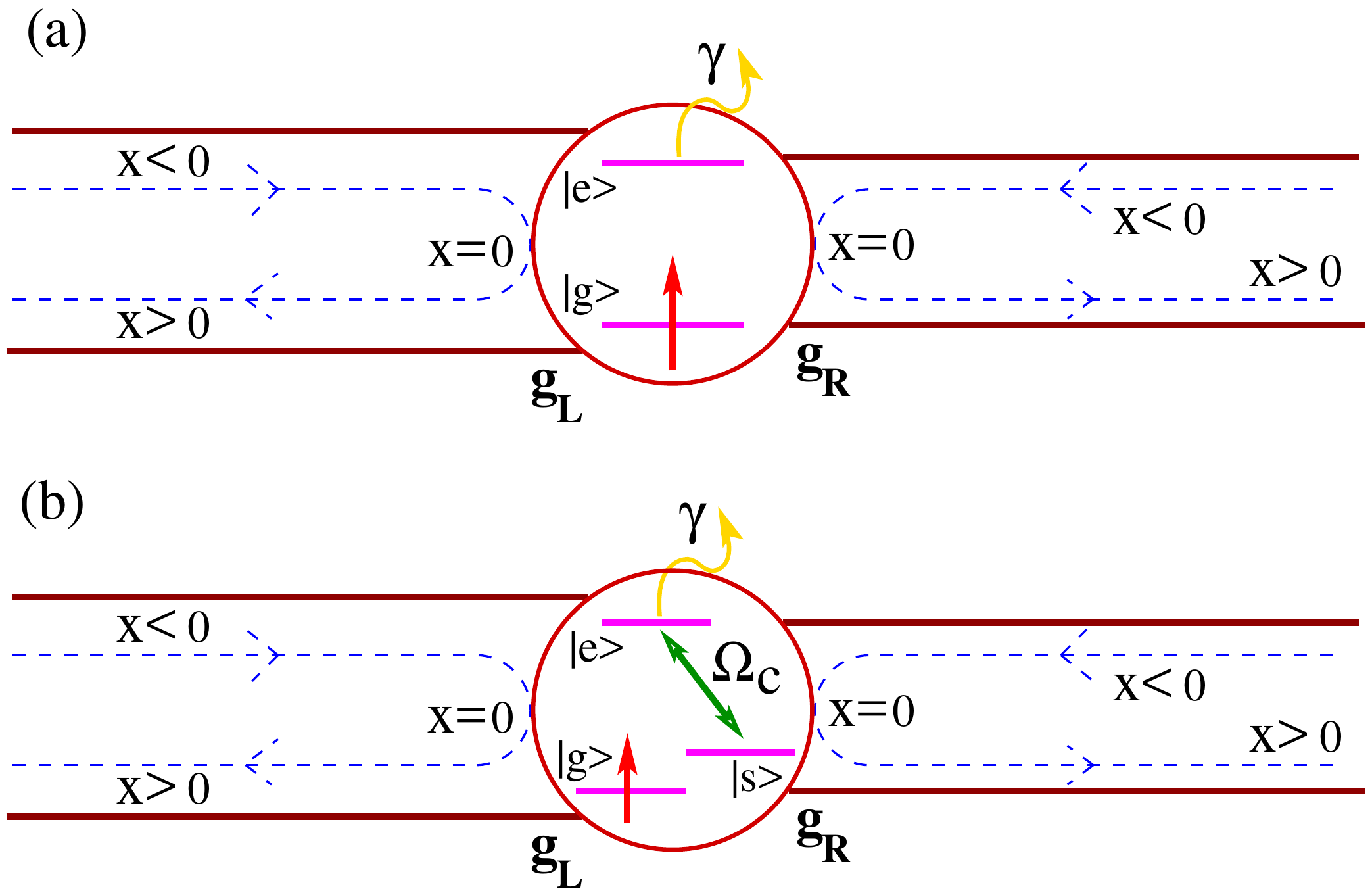}
\caption{Models of nonlinear optical isolator. (a) A two-level atom and (b) a driven $\Lambda$-type three-level atom are coupled asymmetrically $(g_L \ne g_R)$ to light inside open waveguides. A real-space description of the propagating photons is shown by dashed lines where the photons at $x<0$ and $x>0$ represent respectively incoming and scattered parts on each side of the atoms and the photons at $x=0$ are coupled to the atomic transition between $|g\ra$ and $|e\ra$.}
\label{cartoon}
\end{figure}

The proposed all-optical diode can be implemented in experiments with superconducting transmission lines coupled to an artificial atom, such as superconducting qubits \cite{Astafiev10a,Hoi11}, or line-defects in photonic crystals coupled to quantum dots \cite{Lodahl13}. However, some significant modifications in the original calculation are required for an adequate description of these experimental systems. These are (i) incident light in coherent states instead of in Fock states and (ii) incorporation of pure-dephasing and nonradiative decay of the atom either of which is inevitable in such physical systems. In this paper, we address these tasks for two prototypical models of nonlinear optical isolators \cite{RoyPRB2010} using quantum Langevin equations and Green's function method \cite{DharSen2006, DharRoy2006} which is a bit similar \footnote{In contrast to the input-output theory, we do not define input and output operators for the quantum Langevin equations and Green's function method. The latter is easier to implement and has been applied to study nonequilibrium electrical \cite{DharSen2006} and thermal \cite{DharRoy2006} transport in metals, insulators, and superconductors \cite{RoyPRB2012}.} to the popular input-output theory or Heisenberg-Langevin equations approach \cite{Walls08, Gardiner85, Fan10, Koshino12, Peropadre13, Xu15, Fang15, Caneva15, Combes16}. The studied models are (a) a 2LA  being asymmetrically coupled to light inside 1D waveguides and (b) a driven $\Lambda$-type three-level atom (3LA) asymmetrically coupled to light inside 1D waveguides (see Fig.~\ref{cartoon}).

Optical nonlinearity in model (b) is much higher than in model (a) \cite{Roy11}. Here we show that the NR in transmission of light is greater in model (b) than model (a) due to stronger optical nonlinearity. The NR in transmission depends nonmonotonically on the intensity of incoming light and asymmetry in the coupling. We calculate the critical intensity for which NR is maximum and also find a dependence of this maximum NR on asymmetry in the coupling in model (a). To our surprise, we find that while incoherent scattering has a larger contribution in NR at a higher intensity, it is mainly due to coherent elastic scattering of the incident light at a lower power. Finally, we show that NR of an incident light can be improved in the presence of an additional weak light at the opposite port.

The rest of this paper is organized as follows. In Sec.~\ref{sec2LA} we present the theoretical model of an asymmetrically coupled 2LA and derive analytical formulae for the NR and power spectrum of the transmitted light in this model. We compare the NR in transmitted light through a driven 3LA to that through a 2LA in Sec.~\ref{sec3LA}. Sec.~\ref{conc} provides some conclusions and perspectives of our study.

\section{Asymmetrically coupled two-level atom}
\label{sec2LA}
We first consider a 2LA with a transition frequency $\om_e$ between the ground level $|g\ra$ and excited level $|e\ra$. The 2LA is direct-coupled to light inside open waveguides at the left and right side of it with coupling strength $g_{L}$ and $g_{R}$ respectively.  The Hamiltonian of the full system is
\bea
&&\f{\mathcal{H}_{2LA}}{\hbar}=\om_e \sigma^{\dg}\sigma+ \int_{-\infty}^{\infty}dk \big[ v_gk (a^{\dg}_ka_k+b^{\dg}_kb_k+c^{\dg}_kc_k+d^{\dg}_kd_k)\nn\\
&&+\big(\sigma^{\dg}(g_{L}a_k+g_{R}b_k+\gamma d_k)+{\rm h.c.}\big)+\lambda \sigma^{\dg}\sigma(c^{\dg}_k+c_k)\big],\label{Ham}
\eea
where we assume a linear energy-momentum dispersion for different photon modes with a group velocity $v_g$ and write light-matter interactions in linear form within the rotating-wave approximation. Here $\sigma^{\dg}(\equiv |e\ra \la g|)$ and $\sigma(\equiv |g\ra \la e|)$ are respectively the raising and lowering operator of the 2LA, and $a^{\dg}_k,~b^{\dg}_k$ create a photon with wave number $k$ respectively at the left and right side of the 2LA. The operators $c^{\dg}_k,~d^{\dg}_k$ respectively denote creation of excitations related to pure-dephasing (dominant in superconducting circuits) and nonradiative decay. The couplings $\lambda$ and $\gamma$ control the strength of pure-dephasing and nonradiative decay. All the couplings are taken to be constant over photon frequency near $\om_e$; this is known as the Markov approximation causing the photon fields to behave as memoryless baths. We also consider here that the couplings are turned on at $t=t_0$ when light beams are shined on the 2LA. 

We start the calculation by writing the Heisenberg equations of motion for operators $a_k,b_k,c_k,d_k,\sigma$ and $\sigma^{\dg}\sigma$ appearing in the Hamiltonian in Eq.~\ref{Ham}. 
These equations for the photon operators $a_k,b_k,c_k,d_k$ are first-order linear inhomogeneous differential equations which we solve formally for some initial condition at $t_0$. The initial condition of photon operators indicates a direction of incoming photons. We get time-evolution of the photon operators, for example, $a_k(t)$ with an initial condition $a_k(t_0)$ as 
\bea
a_k(t)&=&G_k(t-t_0)a_k(t_0)-ig_L \int_{t_0}^{t} dt'G_k(t-t') \sigma(t'),\label{HEsol1}
\eea
with $G_k(\tau)=e^{-iv_gk\tau}$, and similarly for $b_k(t)$, $c_k(t)$, $d_k(t)$. Plugging these solutions of the photon operators in the Heisenberg equations of the  atomic operators $\sigma$ and $\sigma^{\dg}\sigma$, we find the following equations:
\bea
\f{d\sigma}{dt}&=&-i(\om_e-i\Gamma_t)\sigma-i(1-2\sigma^{\dg}\sigma)\eta_d(t)\nn\\&&-i\lambda (\sigma \eta_c(t)+ \eta_c^{\dg}(t)\sigma),\label{sigma3}\\
\f{d\sigma^{\dg}\sigma}{dt}&=&-2\Gamma_d\sigma^{\dg}\sigma+i\eta_d^{\dg}(t)\sigma-i\sigma^{\dg}\eta_d(t),\label{sigma4}
\eea
where we identify $\eta_d(t)=\int_{-\infty}^{\infty}dk~G_k(t-t_0)(g_La_k(t_0)+g_Rb_k(t_0)+\gamma d_k(t_0))$ and $\eta_c(t)=\int_{-\infty}^{\infty}dk~G_k(t-t_0)c_k(t_0)$ as noises whose properties are determined by the initial condition of the photon fields at $t=t_0$. The rates $\Gamma_d=\Gamma_L+\Gamma_R+\Gamma_{\gamma}$ and $\Gamma_t=\Gamma_d+\Gamma_{\lambda}$ with $\Gamma_L=\pi g_L^2/v_g,~\Gamma_R=\pi g_R^2/v_g,~\Gamma_{\gamma}=\pi\gamma^2/v_g, \Gamma_{\lambda}=\pi\lambda^2/v_g$ denote dissipation and dephasing of the 2LA. The Eqs.~\ref{sigma3},\ref{sigma4} are in the form of quantum Langevin equations, being nonlinear differential equations of operators with multiplicative noises.

The transmission and reflection coefficientss of photons are calculated using a continuity equation, \cite{DharSen2006, DharRoy2006}
\bea
\f{d \sigma^{\dg}\sigma}{dt}+\triangledown.j_p=0,\label{cont}
\eea
where $j_p$ is an operator for photon current. For an incident light from the left of the 2LA, we write $\triangledown.j_p=j_{pb}+j_{pd}-j_{pa}$, where $j_{pa}$ and $j_{pb}$ are photon current respectively at the left and right side of the 2LA, and $j_{pd}$ is current of nonradiative decay. We find these current operators by plugging the Heisenberg equation for $\sigma^{\dg}\sigma$ in Eq.~\ref{cont}:
\bea
j_{pa}(t)&=&ig_L\int_{-\infty}^{\infty}dk (a_k^{\dg}(t)\sigma(t)-\sigma^{\dg}(t)a_k(t)),\label{curr1}\\
j_{pb}(t)&=&-ig_R\int_{-\infty}^{\infty}dk (b_k^{\dg}(t)\sigma(t)-\sigma^{\dg}(t)b_k(t)),\label{curr2} \\
j_{pd}(t)&=&-i\gamma\int_{-\infty}^{\infty}dk (d_k^{\dg}(t)\sigma(t)-\sigma^{\dg}(t)d_k(t)).\label{curr3}
\eea
At steady-state, $\f{d \sigma^{\dg}\sigma}{dt}=0$ which results in $j_{pa}=j_{pb}+j_{pd}$. The transmission and reflection coefficients of light are calculated from $j_{pb}/(v_gI_{\rm in})$ and $1-j_{pa}/(v_gI_{\rm in})$ respectively where $I_{\rm in}$ is the intensity (total number of photons) of the incident light per unit quantization length.

We here consider two different initial conditions of incoming light: (i) a single light beam from one side of the 2LA in Subsec.~\ref{singleB} and (ii) two light beams from opposite sides of the 2LA in Subsec.~\ref{doubleB}. 

\subsection{Single light beam}
\label{singleB}
First, we consider a single input light in a coherent state $|E_p,\om_p\ra$ with a frequency $\om_p$ and an amplitude $E_p$. We take everywhere the amplitude of light to be real for simplicity. For an input light from the left of the 2LA, we have $a_k(t_0)|E_p,\om_p\ra=E_p\delta (v_gk-\om_p)|E_p,\om_p\ra$ and $b_k(t_0)|E_p,\om_p\ra=c_k(t_0)|E_p,\om_p\ra=d_k(t_0)|E_p,\om_p\ra=0$. Thus, we get 
\bea
I_{\rm in}=\la E_p,\om_p|\int dka^{\dg}_k(t_0)a_k(t_0)|E_p,\om_p\ra/\mathcal{L}=\f{E_p^2}{2\pi v_g^2},\label{intensity}
\eea
where we use $\delta(k=0)=\mathcal{L}/(2\pi)$ and $\mathcal{L}$ is the quantization length.

We apply the above properties of coherent state to solve the nonlinear operator  Eqs.~\ref{sigma3},\ref{sigma4}. By performing expectation of these operator Eqs.~\ref{sigma3},\ref{sigma4} in the initial state $|E_p,\om_p\ra$, we convert the noise operators into c-numbers. We define \cite{Koshino12}
\bea
\mathcal{S}_1(t)&=&\la E_p,\om_p|\sigma(t)| E_p,\om_p\ra e^{i\om_p(t-t_0)},\label{s1}\\
\mathcal{S}_2(t)&=& \la E_p,\om_p|\sigma^{\dg}(t)\sigma(t)| E_p,\om_p\ra,\label{s2}
\eea
and $\mathcal{S}_1^{*}(t)=(\mathcal{S}_1(t))^*$ which satisfy a closed set of linear coupled differential equations obtained from Eqs.~\ref{sigma3},\ref{sigma4}. We write these equations in a compact manner by introducing vectors $\boldsymbol{\mathcal{S}}=(\mathcal{S}_1(t),\mathcal{S}_1^*(t),\mathcal{S}_2(t))^{T}$ and $\boldsymbol{\Omega}=(-i\Omega_L, i\Omega_L, 0)^{T}$:
\bea
\f{d\boldsymbol{\mathcal{S}}}{dt}=\left( \begin{array}{ccc} i\delta \omega_p-\Gamma_t  & 0 & 2i\Omega_L\\ 0 & -i\delta \omega_p-\Gamma_t & -2i\Omega_L\\ i\Omega_L & -i\Omega_L & -2\Gamma_d \end{array}\right)\boldsymbol{\mathcal{S}}+\boldsymbol{\Omega},\label{sigma5}
\eea
with detuning $\delta \om_p=\om_p-\om_e$ and Rabi frequency $\Omega_L=g_LE_p/v_g$ for an incident light from the left of the 2LA. The Eq.~\ref{sigma5} for such non-operator variables can be solved for an initial condition, e.g., $\mathcal{S}_1(t=t_0)=\mathcal{S}_1^{*}(t=t_0)=\mathcal{S}_2(t=t_0)=0$ which indicates the 2LA in the ground state before shining a light on it. 
The long-time steady-state behavior of the system is independent of the initial condition for the 2LA. The  steady-state solutions are obtained by setting $\f{d\boldsymbol{\mathcal{S}}}{dt}=0$. These are 
\bea
\mathcal{S}_1(t \to \infty)&=&\mathcal{S}_1(\infty)=\f{-i\Omega_L(i\delta \om_p+\Gamma_t)\Gamma_d}{\Lambda_L},\\
\mathcal{S}_2(t \to \infty)&=&\mathcal{S}_2(\infty)=\f{\Gamma_t\Omega_L^2}{\Lambda_L},
\eea
where $\Lambda_L=\Xi+2\Gamma_t\Omega_L^2$ with $\Xi=\Gamma_d(\Gamma_t^2+\delta \om_p^2)$. 

Using the above solution of $\mathcal{S}_1(\infty)$ and  $\mathcal{S}_2(\infty)$ we evaluate expectation value of the steady-state current operators, $j_{pa},j_{pb},j_{pd}$ in the initial state $|E_p,\om_p\ra$. We denote $\la E_p,\om_p|j_{pa}|E_p,\om_p\ra$ by $\la j_{pa} \ra$ and so forth.
\bea
\la j_{pa} \ra &=& -2(\Omega_L{\rm Im}[\mathcal{S}_1(\infty)]+\Gamma_L\mathcal{S}_2(\infty))\nn\\&=&\f{2\Omega_L^2\Gamma_t(\Gamma_R+\Gamma_{\gamma})}{\Lambda_L},\label{curra}\\
\la j_{pb} \ra &=& 2\Gamma_R\mathcal{S}_2(\infty)=\f{2\Omega_L^2\Gamma_t\Gamma_R}{\Lambda_L},\label{currb}\\
\la j_{pd} \ra &=& 2\Gamma_{\gamma}\mathcal{S}_2(\infty)=\f{2\Omega_L^2\Gamma_t\Gamma_{\gamma}}{\Lambda_L}.\label{currd}
\eea
Indeed $\la j_{pa} \ra=\la j_{pb} \ra + \la j_{pd} \ra$ in the steady-state. The transmission coefficient of light from left to right side of the 2LA is $\mathcal{T}_{LR}=\la j_{pb} \ra /(v_gI_{\rm in})=4\Gamma_t\Gamma_L\Gamma_R/\Lambda_L$. $\mathcal{T}_{LR}$ depends asymmetrically on $\Gamma_L$ and $\Gamma_R$ due to the term $2\Gamma_t\Omega_L^2$ in the denominator $\Lambda_L$. The term $2\Gamma_t\Omega_L^2$ is related to the intensity of incident light from the left. Transmission coefficient $\mathcal{T}_{RL}$ from right to left of the 2LA is found by exchanging $\Gamma_L$ and $\Gamma_R$ in $\mathcal{T}_{LR}$. Thus, a difference in light transmission under reversal of incident light's direction is
\bea
\Delta \mathcal{T}=\mathcal{T}_{LR}-\mathcal{T}_{RL}=4\Gamma_t\Gamma_L\Gamma_R\left(\f{1}{\Lambda_L}- \f{1}{\Lambda_R}\right),\label{rect1}
\eea
with $\Lambda_R=\Xi+2\Gamma_t\Omega_R^2$. The transmission difference $\Delta \mathcal{T}$ is a measure of NR in this system. It vanishes when $g_L=g_R$. It also vanishes in the single-photon limit of the incident light \cite{RoyPRB2010} for $E_p \to 0$ when the terms $2\Gamma_t\Omega_L^2$ in $\Lambda_L$  and $2\Gamma_t\Omega_R^2$ in $\Lambda_R$ are dropped. Therefore, both the asymmetry in coupling and optical nonlinearity at higher light intensity are essential for nonreciprocal transmission of light in the current system. 

\begin{figure*}
\includegraphics[width=0.99\linewidth]{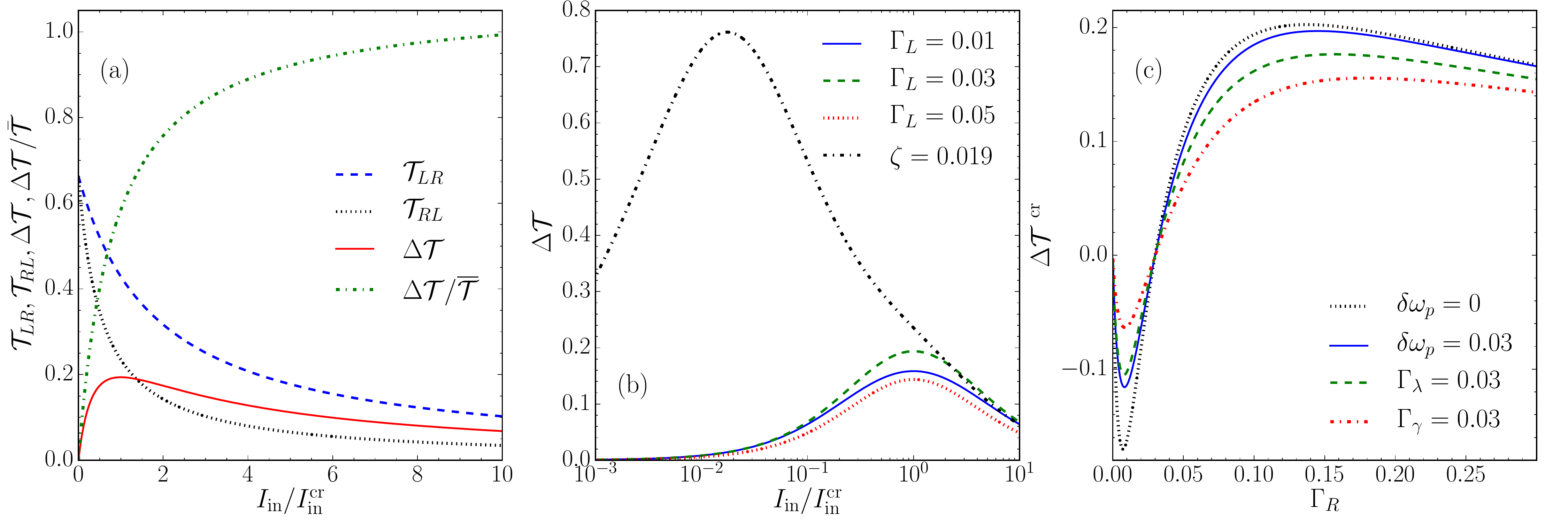}
\caption{Features of nonreciprocal light propagation through an asymmetrically coupled two-level atom. (a) Transmission coefficients $\mathcal{T}_{LR},\mathcal{T}_{RL}$, a difference in transmission $\Delta \mathcal{T}$ and a normalized difference in transmission $\Delta \mathcal{T}/\bar{\mathcal{T}}$ with scaled intensity $I_{\rm in}/I^{\rm cr}_{\rm in}$ of incident light. (b) $\Delta \mathcal{T}$ vs. $I_{\rm in}/I^{\rm cr}_{\rm in}$ at $\Gamma_R=0.1$ and three different $\Gamma_L$. The black dash-dotted curve is for $\Gamma_L=0.03$ and $\Gamma_R=0.1$ in the presence of an additional backward light of intensity $I_{\rm b}$ with $\zeta=I_{\rm b}/I_{\rm in}^{\rm cr}=0.018$. (c) A difference in transmission $\Delta \mathcal{T}^{\rm cr}$ at the critical intensity $I^{\rm cr}_{\rm in}$ with $\Gamma_R$ for a fixed $\Gamma_L=0.03$ and different values of $\Gamma_{\gamma},\Gamma_{\lambda}$ and $\delta \omega_p$. In all the plots $\Gamma_L=0.03,\Gamma_R=0.1,\Gamma_{\gamma}=\Gamma_{\lambda}=0.003$ and $\delta \omega_p=0$ if they are not explicitly mentioned. The rates $\Gamma_L,\Gamma_R,\Gamma_{\gamma},\Gamma_{\lambda}$ and $\delta \omega_p$ are in units of $\om_e$ and $v_g=1$.}
\label{rec1}
\end{figure*}

We plot lineshape of $\mathcal{T}_{LR},\mathcal{T}_{RL}$ and $\Delta \mathcal{T}$ with a scaled intensity of incident light in Fig.~\ref{rec1}(a) for $\Gamma_L/\om_e=0.03$ and $\Gamma_R/\om_e=0.1$. Both  $\mathcal{T}_{LR},\mathcal{T}_{RL}$ fall monotonically with increasing $I_{\rm in}$ due to photon-blockade in a direct-coupled system.
A 2LA is saturated by a single photon; therefore it acts as a nonlinear medium for two or multiple photons. However, the strength of such optical nonlinearity by a single 2LA is expected to fall above a critical photon number as multiple photons can not simultaneously interact with a 2LA. Consequently, nonreciprocity in light transmission through a 2LA would decrease above a critical intensity of the incident light. Indeed we find from Eq.~\ref{rect1} that $\Delta \mathcal{T}$ increases with $I_{\rm in}$ up to a critical value $I^{\rm cr}_{\rm in}=\Xi/(4v_g\Gamma_t\sqrt{\Gamma_L\Gamma_R})$ before falling monotonically with a further increase in $I_{\rm in}$ \cite{FratiniPRA2016}. The nonmonotonic nature of $\Delta \mathcal{T}$ with increasing $I_{\rm in}$ is shown in Fig.~\ref{rec1}(a,b) for $\Gamma_R/\om_e=0.1$ and different $\Gamma_L/\om_e=0.01,0.03,0.05$. In Fig.~\ref{rec1}(a) we also plot a scaled transmission difference $\Delta \mathcal{T}/\bar{\mathcal{T}} \equiv 2\Delta \mathcal{T}/(\mathcal{T}_{LR}+\mathcal{T}_{RL})$ which increases monotonically with increasing $I_{\rm in}$ before saturating at large $I_{\rm in}$. Interestingly, $\Delta \mathcal{T}/\bar{\mathcal{T}}$ at high $I_{\rm in}$ shows a large value where light transmission through the 2LA itself is a bit small.

For fixed $\Gamma_L$ and $\Gamma_R$, the maximum NR, $\Delta \mathcal{T}^{\rm cr}$ is achieved at $I_{\rm in}=I^{\rm cr}_{\rm in}$,
\bea
\Delta \mathcal{T}^{\rm cr}=\f{4\Gamma_t\Gamma_L\Gamma_R(\Gamma_R-\Gamma_L)}{\Xi(\Gamma_R+\Gamma_L+2\sqrt{\Gamma_L\Gamma_R})}. \label{rect2}
\eea 
The dependence of $\Delta \mathcal{T}^{\rm cr}$ on asymmetry in the coupling is nontrivial which can be seen by plotting $\Delta \mathcal{T}^{\rm cr}$ with $\Gamma_L$ and $\Gamma_R$. We plot $\Delta \mathcal{T}^{\rm cr}$ with $\Gamma_R/\om_e$ in Fig.~\ref{rec1}(c) for a fixed $\Gamma_L/\om_e(=0.03)$. $\Delta \mathcal{T}^{\rm cr}$ changes nonmonotonically with $\Gamma_R$ and its sign switches across $\Gamma_R=\Gamma_L$. For a fixed $\Gamma_L$, $\Delta \mathcal{T}^{\rm cr}$ becomes extreme at two values of $\Gamma_R$ -- one value is smaller than $\Gamma_L$, and another is larger than $\Gamma_L$. In Fig.~\ref{rec1}(c) we also show how detuning $\delta \om_p$, pure dephasing $\Gamma_{\lambda}$, and nonradiative decay $\Gamma_{\gamma}$ affect NR. We find while the magnitude $\Delta \mathcal{T}^{\rm cr}$ decreases with an increasing value of $\Gamma_{\gamma}$ or $\Gamma_{\lambda}$ or $\delta \om_p$,  $\Gamma_{\gamma}$ has a relatively higher influence on NR than those due to $\delta \om_p$ or $\Gamma_{\lambda}$. Here we point out that higher ratios of $\Gamma_L/\Gamma_{\gamma}, \Gamma_R/\Gamma_{\gamma}, \Gamma_L/\Gamma_{\lambda}, \Gamma_R/\Gamma_{\lambda}$ can create a higher NR even for smaller values of $\Gamma_L$ and $\Gamma_R$. We also notice that the maximum value of $\Delta \mathcal{T}^{\rm cr}$ for our studied parameters in this model is around $0.2$ which is somewhat small for practical applications.

To have a better understanding of the underlying physical mechanism of this NR, we now investigate a role of coherent and incoherent scattering of light in NR. Coherently scattered light has a constant phase relation with the incident light and can be detected using phase-sensitive homodyne-type measurements. We evaluate power spectrum of the transmitted light to find coherent and incoherent parts in it. For this, we introduce a real-space description of the photon operators at position $x \in [-\infty,\infty]$ of both sides of the 2LA (check Fig.~\ref{cartoon}). We define photon operator as $a_x(t)=\int_{-\infty}^{\infty} dk e^{ikx}a_k(t)/\sqrt{2\pi}$ and $b_x(t)=\int_{-\infty}^{\infty} dk e^{ikx}b_k(t)/\sqrt{2\pi}$ where the operators at $x<0$ and $x>0$ represent respectively incident and scattered photons on each side of the 2LA and the photons at $x=0$ are coupled to the 2LA.

For an incident light from left of the 2LA, the power spectrum of transmitted light at long-time steady-state is defined as
\bea
P_{\rm tr}(\om)={\rm Re}\int_0^{\infty}\f{d\tau}{\pi}e^{i\om \tau}\la b_x^{\dg}(t)b_{x}(t+\tau)\ra,\label{power1} 
\eea
where we take $x >0, t\gg t_0$ and the expectation $\la..\ra$ is again performed in the initial state $|E_p,\om_p\ra$. An expression like Eq.~\ref{power1} for $a_x(t)$ at $x <0, t<t_0$ would give a power spectrum of the incident light, $P_{\rm in}(\om)=E_p^2\delta(\om-\om_p)/(2\pi v_g^2)$. Thus, total incident power, $\int d\om P_{\rm in}(\om)=I_{\rm in}$. To calculate $P_{\rm tr}(\om)$, we first apply a formal solution of the Heisenberg equation for $b_k(t)$ like that of Eq.~\ref{HEsol1}, and rewrite  $P_{\rm tr}(\om)$ using input fields and atomic operators. Applying $b_k(t_0)|E_p,\om_p\ra=0$, we find
\bea
P_{\rm tr}(\om)=\f{2\Gamma_R}{\pi v_g}{\rm Re}\int_0^{\infty}d\tau e^{i\om \tau} \la \sigma^{\dg}(t)\sigma(t+\tau)\ra.\nn
\eea
Thus, we now need to calculate a two-time correlation of atomic operators $\la \sigma^{\dg}(t)\sigma(t+\tau)\ra$ to proceed further. So we define three new correlators \cite{Koshino12}: $\mathcal{S}_3(\tau)=\la \sigma^{\dg}(t)\sigma(t+\tau)\ra e^{i\om_p\tau}$, $\mathcal{S}_4(\tau)=\la \sigma^{\dg}(t)\sigma^{\dg}(t+\tau)\ra e^{-i\om_p(2(t-t_0)+\tau)}$, and $\mathcal{S}_5(\tau)=\la \sigma^{\dg}(t)\sigma^{\dg}(t+\tau)\sigma(t+\tau)\ra e^{-i\om_p(t-t_0)}$, which are $t$-independent at long-time steady-state. Notice here, $\int d\omega P_{\rm tr}(\omega)/I_{\rm in}=2\Gamma_R \langle \sigma^{\dg}(t)\sigma(t) \rangle/(v_gI_{\rm in})=\mathcal{T}_{LR}$.

\begin{figure*}
\includegraphics[width=0.99\linewidth]{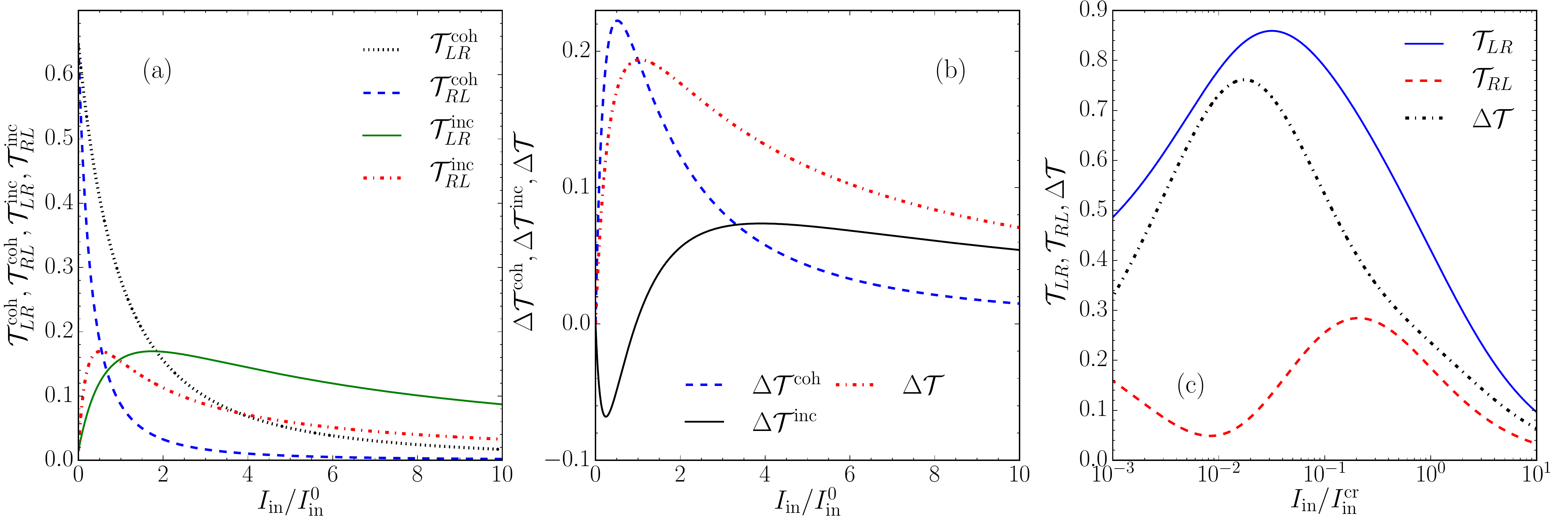}
\caption{Contributions of coherent and incoherent scattering and additional backward light in nonreciprocal light propagation through an asymmetrically coupled two-level atom. (a) A contribution of coherent and incoherent scattering in transmission coefficients $\mathcal{T}_{LR},\mathcal{T}_{RL}$ with scaled intensity $I_{\rm in}/I^{0}_{\rm in}$ of incident light. (b) A contribution of coherent $(\Delta \mathcal{T}^{\rm coh})$ and incoherent $(\Delta \mathcal{T}^{\rm inc})$ scattering in nonreciprocity. (c) $\mathcal{T}_{LR},\mathcal{T}_{RL}$ and $\Delta \mathcal{T}$ vs. $I_{\rm in}/I^{\rm cr}_{\rm in}$ of a forward light in the presence of an additional backward light of intensity $I_{\rm b}$ with $I_{\rm b}/I_{\rm in}^{\rm cr}=0.018$. In all the plots $\Gamma_L=0.03,\Gamma_R=0.1,\Gamma_{\gamma}=\Gamma_{\lambda}=0.003$ and $\delta \omega_p=0$. The rates $\Gamma_L,\Gamma_R,\Gamma_{\gamma},\Gamma_{\lambda}$ and $\delta \omega_p$ are in units of $\om_e$ and $v_g=1$.}
\label{rec1e}
\end{figure*}

It can be shown after some algebra that $\mathcal{S}_3(\tau),~\mathcal{S}_4(\tau)$ and $\mathcal{S}_5(\tau)$ satisfy a set of inhomogeneous differential equations similar to those in Eq.~\ref{sigma5} when $\boldsymbol{\mathcal{S}}$ and $\boldsymbol{\Omega}$ are replaced respectively by $\tilde{\boldsymbol{\mathcal{S}}}=(\mathcal{S}_3(t),\mathcal{S}_4(t),\mathcal{S}_5(t))^{T}$ and $\tilde{\boldsymbol{\Omega}}=(-i\Omega_L\mathcal{S}^*_1(\infty),i\Omega_L\mathcal{S}^*_1(\infty),0)^T$. The initial condition and asymptotic behavior in the limit $\tau \to \infty$  of these correlations are: $\mathcal{S}_3(\tau=0)=\mathcal{S}_2(\infty),~\mathcal{S}_4(\tau=0)=\mathcal{S}_4(\tau=0)=0$, and $\mathcal{S}_3(\tau \to \infty)=|\mathcal{S}_1(\infty)|^2$, $\mathcal{S}_4(\tau \to \infty)=(\mathcal{S}_1^*(\infty))^2$, $\mathcal{S}_5(\tau \to \infty)=\mathcal{S}_1^*(\infty)\mathcal{S}_2(\infty)$.
Using these long-$\tau$ limit, we now define $\delta \mathcal{S}_j(\tau)=\mathcal{S}_j(\tau)-\mathcal{S}_j(\tau \to \infty)$ with $j=3,4,5$ which satisfy a set of homogeneous differential equations similar to the homogeneous part of Eq.~\ref{sigma5}. We solve these coupled differential equations to evaluate the power spectrum $P_{\rm tr}(\om)=P^{\rm coh}_{\rm tr}(\om)+P^{\rm inc}_{\rm tr}(\om)$ where $P^{\rm coh}_{\rm tr}(\om)$ and $P^{\rm inc}_{\rm tr}(\om)$ represent respectively coherent and incoherent parts of the transmitted power. The coherent part of transmitted power is $P^{\rm coh}_{\rm tr}(\omega)=(2 \Gamma_R/v_g)|\mathcal{S}_1(\infty)|^2\delta(\omega-\omega_p)$, and the incoherent part  of transmitted power is a $\omega$-dependent long expression which we do not show here.

For an incoming light from the left of the 2LA, we calculate coherent and incoherent parts of the total transmitted power by taking integration over $\om$ of $P^{\rm coh}_{\rm tr}(\om)$ and $P^{\rm inc}_{\rm tr}(\om)$ respectively. We define $\int d\omega P^{\rm coh}_{\rm tr}(\omega)/I_{\rm in}=\mathcal{T}^{\rm coh}_{LR}$ and $\int d\omega P^{\rm inc}_{\rm tr}(\omega)/I_{\rm in}=\mathcal{T}^{\rm inc}_{LR}$ for an incoming light from the left. By switching $\Gamma_L$ and $\Gamma_R$, we find coherent and incoherent parts of the total transmitted power for an incoming light from the right of the 2LA. The Fig.~\ref{rec1e}(a) depicts the contribution of coherent and incoherent scattering in transmission coefficient of light from the left or the right side of the 2LA. It shows that the transmission of light through a 2LA is entirely due to coherent scattering at low light power (at single-photon limit). The incoherent scattering has a larger contribution in transmission at higher light power. One important feature in Fig.~\ref{rec1e}(a) is that not only the incoherent  scattering but the coherent scattering is also sensitive to the asymmetry in the coupling at a finite intensity. It can be understood following nonlinear optical processes in light propagation through a nonlinear medium. The third-order nonlinear optical processes lead to an intensity-dependent nonlinear contribution to the refractive index experienced by a light at incident frequency $\omega_p$ \cite{Boyd2008}. This nonlinear refractive index is also spatially asymmetric due to the asymmetric coupling and is responsible for nonreciprocity in the coherently scattered light. 

A difference in total coherent and incoherent transmitted power under reversal of incident light's direction and  after scaling by total incoming power $I_{\rm in}$ are:
\bea
\Delta \mathcal{T}^{\rm coh}\equiv \mathcal{T}^{\rm coh}_{LR}-\mathcal{T}^{\rm coh}_{RL}&=&4\Gamma_L\Gamma_R\Gamma_d\Xi(\Lambda_L^{-2}-\Lambda_R^{-2}),\label{cohpower}\\
\Delta \mathcal{T}^{\rm inc}\equiv \mathcal{T}^{\rm inc}_{LR}-\mathcal{T}^{\rm inc}_{RL}&=&4\Gamma_L\Gamma_R\big(\Gamma_{\lambda}\Xi(\Lambda_L^{-2}-\Lambda_R^{-2})\nn\\&+&2\Gamma_t^2(\Omega_L^2\Lambda_L^{-2}-\Omega_R^2\Lambda_R^{-2})\big).\label{incpower}
\eea
As expected, we get $\Delta \mathcal{T}^{\rm coh}+ \Delta \mathcal{T}^{\rm inc}=\Delta \mathcal{T}$ of Eq.~\ref{rect1}. $\Delta \mathcal{T}^{\rm inc}$ goes through zero at a finite incident intensity $I^0_{\rm in}$ where 
\bea
I^0_{\rm in}&=&\f{1}{2v_g}\left(-\rho_1+\sqrt{\rho_1^2-\rho_2}\right)~~{\rm with}\nn\\ \rho_1&=&\f{\Xi\Gamma_{\lambda}(\Gamma_R+\Gamma_L)}{4\Gamma_t^2\Gamma_R\Gamma_L},\rho_2=\f{\Xi^2(\Gamma_{\lambda}^2-\Gamma_d^2)}{4\Gamma_t^4\Gamma_R\Gamma_L}.
\eea
We plot $\Delta \mathcal{T}^{\rm coh},\Delta \mathcal{T}^{\rm inc}$ and $\Delta \mathcal{T}$ with a scaled intensity $I_{\rm in}/I^0_{\rm in}$ in Fig.~\ref{rec1e}(b). The Fig.~\ref{rec1e}(b) shows that $\Delta \mathcal{T}^{\rm coh}$ and $\Delta \mathcal{T}^{\rm inc}$ have opposite sign for $I_{\rm in}<I^0_{\rm in}$ where they act against each other to reduce $\Delta \mathcal{T}$. Interestingly, the main contribution to NR at lower light power comes from the coherently scattered light at an incident frequency. It indicates that the mixing of incident photon modes is not essential for NR \cite{RoyPRB2010}. NR at a higher light power is mainly due to incoherent scattering. As we have discussed above, both the coherent and incoherent scattering of incident light experience an intensity-dependent and spatially asymmetric refractive index in this system. While the coherent scattering has a larger contribution in light propagation at a lower intensity, the incoherent scattering has a significant contribution to a higher power. However, the total light transmission, as well as NR, fall rapidly with increasing intensity due to photon-blockade in this system. Therefore, the critical intensity $I_{\rm in}^{\rm cr}$ with the maximum nonreciprocity occurs at a relatively lower power when the coherent scattering has the main contribution to NR.  

\subsection{Two light beams}
\label{doubleB}
Next, we consider the presence of an additional small-amplitude backward light along with a large-amplitude forward light. Shi {\it et. al.} \cite{Shi2015} have recently studied this situation to show a decline in NR due to dynamic reciprocity in a nonlinear optical isolator for an additional backward light whose spectral band does not overlap with the forward light. We here investigate the other case when the spectral band of backward light overlaps with the forward light. In this case, we have initial condition $a_k(t_0)|\phi\ra=E_p\delta (v_gk-\om_p)|\phi\ra$, $b_k(t_0)|\phi\ra=E_b\delta (v_gk-\om_b)|\phi\ra$ and $c_k(t_0)|\phi\ra=d_k(t_0)|\phi\ra=0$ for an initial state $|\phi\ra=|E_p,\om_p\ra \otimes |E_b,\om_b\ra$.  $|\phi\ra$ is a product of the states of forward and backward light at the left and right side of the atom with respective frequency $\om_p,\om_b$ and amplitude $E_p,E_b$. We are mostly interested in the regime when $E_b<E_p$ and we take $\om_b=\om_p$ for an overlap of the spectrum of monochromatic lights. We next find steady-state variables $\mathcal{S}_1(t)$ and $\mathcal{S}_2(t)$ in Eqs.~\ref{s1},\ref{s2} for the initial state $|\phi\ra$ and use them to calculate corresponding $\la j_{pa} \ra$ and $\la j_{pb} \ra$.

The total transmission coefficient $\mathcal{T}_{LR}$ of light from left to right side of the 2LA is found by dividing $\la j_{pb} \ra/v_g$ by total intensity $(I_{\rm in}+I_{\rm b})$ of forward and backward light where $I_{\rm b}=E^2_b/(2\pi v_g^2)$ is an intensity of backward light.  As before, we get transmission coefficient $\mathcal{T}_{RL}$ for a forward light from right to left of the 2LA by exchanging $\Gamma_L$ and $\Gamma_R$ in the above $\mathcal{T}_{LR}$. In Fig.~\ref{rec1e}(c), we show $\mathcal{T}_{LR}$ and $\mathcal{T}_{RL}$ with $I_{\rm in}/I^{\rm cr}_{\rm in}$ of a forward light in the presence of an additional backward light of intensity $I_{\rm b}$ with $\zeta=I_{\rm b}/I_{\rm in}^{\rm cr}=0.018$. At high intensities of the forward light when $I_{\rm in} \gg I_{\rm b}$, both $\mathcal{T}_{LR}$ and $\mathcal{T}_{RL}$ in  Fig.~\ref{rec1e}(c) in the presence of an additional backward light fall with increasing $I_{\rm in}$ as the case in Fig.~\ref{rec1}(a) in the absence of an additional backward light. However, the lineshapes of $\mathcal{T}_{LR}$ and $\mathcal{T}_{RL}$ in  Fig.~\ref{rec1e}(c) differ significantly from those in Fig.~\ref{rec1}(a) for lower intensities of the forward light. These differences in transmission are due to nonlinear optical processes between the two lights at the atom acting as a nonlinear medium. These nonlinear processes are sensitive to intensities, frequencies, and phases of the forward and backward lights. The nonlinear processes are significant when the power of the two lights are comparable, i.e., $I_{\rm in} \approx I_{\rm b}$. Therefore, we find some drastic changes both in $\mathcal{T}_{LR}$ and $\mathcal{T}_{RL}$ due to the presence of an additional backward light when $I_{\rm in}$ is comparable to $I_{\rm b}$.  

We also add the NR, $\Delta \mathcal{T}=\mathcal{T}_{LR}-\mathcal{T}_{RL}$ of the forward light in both Fig.~\ref{rec1}(b) and Fig.~\ref{rec1e}(c). We observe a  substantial enhancement in  $\Delta \mathcal{T}$ at $I_{\rm in}<I^{\rm cr}_{\rm in}$ in the presence of an additional backward light (check Fig.~\ref{rec1}(b)), and  $\Delta \mathcal{T}$ has a peak around $I_{\rm in} \approx I_{\rm b}$. It is due to the amplification of nonlinear optical processes by the  additional backward light when $I_{\rm in} \approx I_{\rm b}$. 
Thus, we have an overall improvement in $\Delta \mathcal{T}$ whenever $I_{\rm in} \sim I_{\rm b}$. However, a practical regime of light powers is $I_{\rm b}<I_{\rm in}$ when the increase NR of a large-amplitude forward light due to a small-amplitude backward light may have some potential application.

\section{Asymmetrically coupled and driven $\Lambda$-type three-level atom}
\label{sec3LA}
In this section, we calculate NR in transmission of a probe light through a direct-coupled and driven 3LA. The $\Lambda$-type 3LA has two lower energy levels $|g\ra$ and $|s\ra$, and an excited level $|e\ra$. A monochromatic laser of frequency $\om_c$ strongly drives the transition between levels $|e\ra$ and $|s\ra$. The frequency $\om_c$ is close to the transition frequency $\om_{es}$ between $|e\ra$ and $|s\ra$. The probe light is asymmetrically coupled to the optical transition between $|g\ra$ to $|e\ra$ [see Fig.~\ref{cartoon}(b)] as in the previous section. The Hamiltonian of the full system becomes static in a frame rotating at the drive frequency $\om_c$. Then, 
\bea
\f{\mathcal{H}_{3LA}}{\hbar}&=&\f{\mathcal{H}_{2LA}}{\hbar}+(\om_e+\Delta_c)\mu^{\dg}\mu+\Omega_c(\mu+\mu^{\dg})\nn\\&+&\int_{-\infty}^{\infty}dk \big(v_gk f_k^{\dg}f_k+\lambda'\mu^{\dg}\mu(f_k^{\dg}+f_k)\big),\label{Ham3LA}
\eea
where the detuning $\Delta_c=\om_c-\om_{es}$, the raising operator $\mu^{\dg}=|s\ra \la e|$ and $\Omega_c$ is the Rabi frequency of driving. The coupling $\lambda'$ controls the strength of pure-dephasing of the level $|s\ra$ to a bath of excitations created by $f_k^{\dg}$. 

Here we consider that a single probe light in a coherent state $|E_p,\om_p\ra$ is shined on the driven 3LA at $t=t_0$. Following the steps in the previous section, we then derive quantum Langevin equations for the operators $\sigma,\sigma^{\dg}\sigma,\mu,\mu^{\dg}\mu$ and $\nu~(\equiv |s\ra \la g|)$ of the driven 3LA by integrating out the photon fields. These are,
\bea
\f{d\sigma}{dt}&=&-i(\om_e-i\Gamma_t)\sigma-i\Omega_c\nu^{\dg}-i(1-2\sigma^{\dg}\sigma-\mu^{\dg}\mu)\eta_d(t)\nn\\&&-i\lambda (\sigma \eta_c(t)+ \eta_c^{\dg}(t)\sigma),\label{sigma3a}\\
\f{d\sigma^{\dg}\sigma}{dt}&=&-2\Gamma_d\sigma^{\dg}\sigma-i\Omega_c(\mu-\mu^{\dg})+i\eta_d^{\dg}(t)\sigma-i\sigma^{\dg}\eta_d(t),\label{sigma4a}\\
\f{d\mu}{dt}&=&-i(\Delta_c-i\Gamma_s)\mu+i\Omega_c(\mu^{\dg}\mu-\sigma^{\dg}\sigma)+i\eta^{\dg}_d(t)\nu^{\dg}\nn\\&&+i\lambda (\mu \eta_c(t)+ \eta_c^{\dg}(t)\mu)-i\lambda'(\mu \eta_f(t)+ \eta_f^{\dg}(t)\mu),\label{mu1}\\
\f{d\mu^{\dg}\mu}{dt}&=&i\Omega_c(\mu-\mu^{\dg}),\label{mu2}\\
\f{d\nu}{dt}&=&i(\om_e+\Delta_c+i\Gamma_{\lambda'})\nu+i\Omega_c\sigma^{\dg}-i\eta_d^{\dg}(t)\mu^{\dg}\nn\\&&-i\lambda'(\nu \eta_f(t)+ \eta_f^{\dg}(t)\nu),\label{nu1}
\eea
where $\eta_d(t),\eta_c(t)$ and $\eta_f(t)=\int_{-\infty}^{\infty}dk~G_k(t-t_0)f_k(t_0)$ are noises due to photon fields, and dissipation and dephasing of the driven 3LA are $\Gamma_d,\Gamma_t,\Gamma_{\lambda'}$ and $\Gamma_s=\Gamma_t+\Gamma_{\lambda'}$ with $\Gamma_{\lambda'}=\pi\lambda'^2/v_g$. Apart from those relations for $|E_p,\om_p\ra$ in Subsec.~\ref{singleB} before Eq.~\ref{intensity}, we also have $f_k(t_0)|E_p,\om_p\ra =0$. 

 Next we take expectation of the above quantum Langevin equations of atomic operators in the initial state $|E_p,\om_p\ra$, and rewrite them using non-operator variables. For this we introduce
\bea
\mathcal{M}_1(t)&=&\la E_p,\om_p|\mu(t)|E_p,\om_p\ra,\\
\mathcal{M}_2(t)&=&\la E_p,\om_p|\mu^{\dg}(t)\mu(t)|E_p,\om_p\ra,\\
\mathcal{N}(t)&=&\la E_p,\om_p|\nu^{\dg}(t)|E_p,\om_p\ra e^{i\om_p(t-t_0)}.
\eea
The non-operator variables $\mathcal{S}_1(t),\mathcal{S}^*_1(t),\mathcal{S}_2(t),\mathcal{M}_1(t)$, $\mathcal{M}^*_1(t),\mathcal{M}_2(t),\mathcal{N}(t)$ and $\mathcal{N}^*(t)$ obey a closed set of linear coupled differential equations which are obtained from Eqs.~\ref{sigma3a}-\ref{nu1}. These equations for the input probe light from the left of the driven 3LA are
\bea
&&\f{d\boldsymbol{\mathcal{M}}}{dt}=\boldsymbol{\mathcal{R}}\boldsymbol{\mathcal{M}}+\boldsymbol{\Omega}_M,~~{\rm where}~~\boldsymbol{\mathcal{R}}=\label{eom3LA}\\
&&\left( \begin{array}{cccccccc} \kappa_1  & 0 & 2i\Omega_L & 0 & 0 & i\Omega_L & -i\Omega_c & 0 \\ 0 & \kappa_1^* & -2i\Omega_L & 0 & 0 & -i\Omega_L & 0 & i\Omega_c \\ i\Omega_L & -i\Omega_L & -2\Gamma_d & -i\Omega_c & i\Omega_c & 0 & 0 & 0 \\ 0 & 0 & -i\Omega_c & \kappa_2^* & 0 & i\Omega_c & i\Omega_L & 0 \\ 0 & 0 & i\Omega_c & 0 & \kappa_2 & -i\Omega_c & 0 & -i\Omega_L \\ 0 & 0 & 0 & i\Omega_c & -i\Omega_c & 0 & 0 & 0 \\ -i\Omega_c & 0 & 0 & i\Omega_L & 0 & 0 & \kappa_3  & 0 \\ 0 & i\Omega_c & 0 & 0 & -i\Omega_L & 0 & 0 & \kappa_3^* \end{array}\right),\nn
\eea
$\boldsymbol{\mathcal{M}}(t)=(\mathcal{S}_1,\mathcal{S}^*_1,\mathcal{S}_2,\mathcal{M}_1,\mathcal{M}^*_1,\mathcal{M}_2,\mathcal{N},\mathcal{N}^*)^{T}$ and $\boldsymbol{\Omega}_M=(-i\Omega_L, i\Omega_L, 0, 0, 0 ,0 ,0 ,0)^{T}$
with $\kappa_1=i\delta \omega_p-\Gamma_t,\kappa_2=i\Delta_c-\Gamma_s,\kappa_3=i(\delta \omega_p-\Delta_c)-\Gamma_{\lambda'}$. As before, we are mainly interested in the long-time steady-state solution of Eq.~\ref{eom3LA}, therefore we set $\f{d\boldsymbol{\mathcal{M}}}{dt}=0$. Thus, we get $\boldsymbol{\mathcal{M}}(t \to \infty)=-\boldsymbol{\mathcal{R}}^{-1}\boldsymbol{\Omega}_M$.

\begin{figure}
\includegraphics[width=0.90\linewidth]{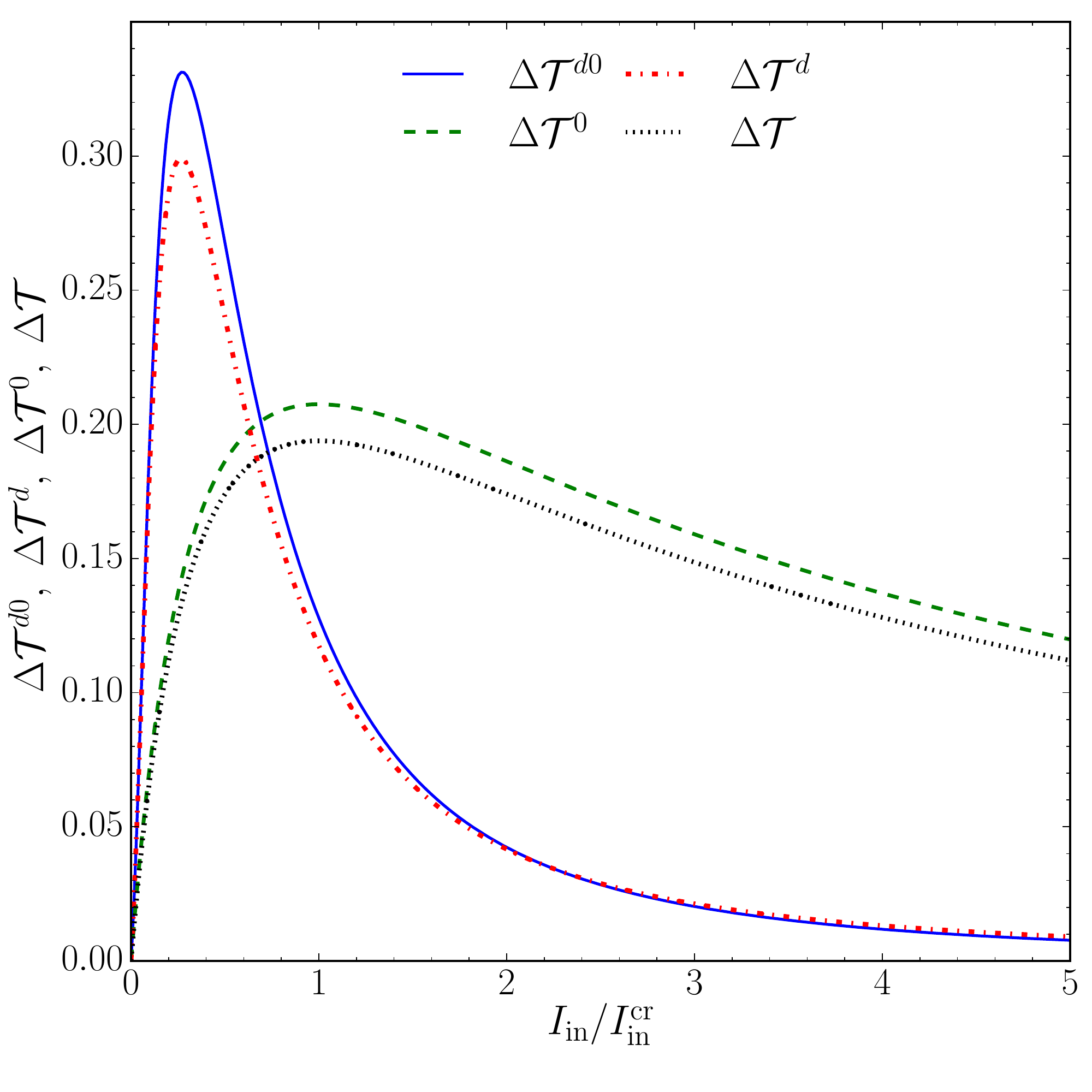}
\caption{A comparison between nonreciprocity in transmission coefficient through a driven $\Lambda$-type three-level atom ($\Delta \mathcal{T}^{d0},\Delta \mathcal{T}^{d}$) and that through a two-level atom ($\Delta \mathcal{T}^{0},\Delta \mathcal{T}$) with a scaled intensity $I_{\rm in}/I^{\rm cr}_{\rm in}$ of an incident probe light. $\Delta \mathcal{T}^{d0},\Delta \mathcal{T}^{0}$ are for lossless atoms, and the rates of loss for $\Delta \mathcal{T}^{d},\Delta \mathcal{T}$ are  $\Gamma_{\gamma}=\Gamma_{\lambda}=0.003$ and $\Gamma_{\lambda'}=0.001$. Here, $\Gamma_L=0.03,\Gamma_R=0.1,\delta \omega_p=0,\Delta_c=0.02$ and $\Omega_c=0.01$. The rates $\Gamma_L,\Gamma_R,\Gamma_{\gamma},\Gamma_{\lambda},\Gamma_{\lambda'},\Omega_c$ and $\delta \omega_p,\Delta_c$ are in units of $\om_e$ and $v_g=1$.}
\label{3LArec1}
\end{figure}

The transmission coefficient $\mathcal{T}^d_{LR}$ of a probe beam from left to right side of the driven 3LA can be found from the ratio of total transmitted and incident power. We get $\mathcal{T}^d_{LR}=2\Gamma_R \mathcal{S}_2(\infty)/(v_gI_{\rm in})$. The expression of $\mathcal{S}_2(\infty)$ is very long in the presence of losses in the driven atom. Therefore, we first consider the lossless case and write $\mathcal{T}^{d}_{LR}$ (called $\mathcal{T}^{d0}_{LR}$) for $\lambda=\gamma=\lambda'=0$. Thus we find
\bea
&&\mathcal{T}^{d0}_{LR}=\label{3LAtrans2}\\&&\f{4\Gamma_L\Gamma_R\Delta^2}{(\Gamma_L+\Gamma_R)^2\Delta^2+(\Omega_c^2-\delta \om_p\Delta)^2+2(\Omega_c^2+\Delta^2)\Omega_L^2+\Omega_L^4},\nn
\eea
with $\Delta=\delta \om_p-\Delta_c$. The transmission $\mathcal{T}^{d0}_{LR}$ vanishes at zero $\Delta$ (i.e., $\delta \om_p=\Delta_c)$ and the corresponding reflection coefficient reaches unity. This is a reminiscent of the electromagnetically induced transparency (EIT) observed in a side-coupled driven 3LA \cite{Roy11, Hoi11}. The transmission and reflection coefficients in a side-coupled atom alter to the reflection and transmission coefficients in a direct-coupled atom \cite{Roy2017}. However, a more important feature in $\mathcal{T}^{d0}_{LR}$ is the emergence of a higher order optical nonlinearity compared to the direct-coupled 2LA. It is due to the $\Omega_L^4$ term in the denominator of $\mathcal{T}^{d0}_{LR}$ which is absent in the transmission through a lossless 2LA, $\mathcal{T}^0_{LR}=4\Gamma_L\Gamma_R/((\Gamma_L+\Gamma_R)^2+\delta \om_p^2+2\Omega_L^2)$.

\begin{figure}
\includegraphics[width=0.90\linewidth]{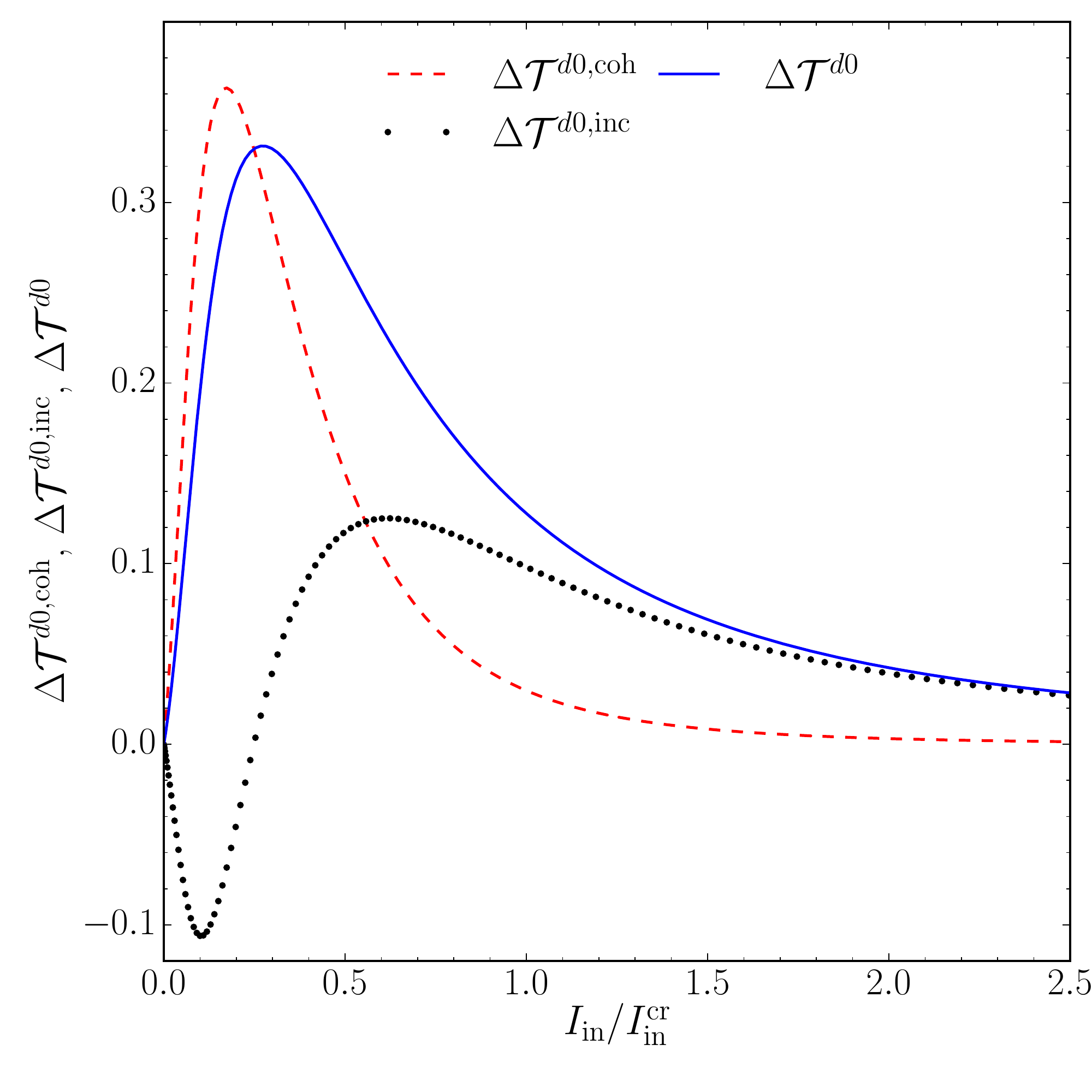}
\caption{A contribution of coherent $(\Delta \mathcal{T}^{d0,\rm coh})$ and incoherent $(\Delta \mathcal{T}^{d0,\rm inc})$ scattering in nonreciprocity in total transmission $(\Delta \mathcal{T}^{d0})$ through the driven $\Lambda$-type three-level atom without loss. Here, $\Gamma_L=0.03,\Gamma_R=0.1,\delta \omega_p=0,\Delta_c=0.02$, $\Omega_c=0.01$ and $\Gamma_{\gamma}=\Gamma_{\lambda}=\Gamma_{\lambda'}=0$. The rates $\Gamma_L,\Gamma_R,\Gamma_{\gamma},\Gamma_{\lambda},\Gamma_{\lambda'},\Omega_c$ and $\delta \omega_p,\Delta_c$ are in units of $\om_e$ and $v_g=1$.}
\label{3LArec2}
\end{figure}

The higher nonlinearity in the driven 3LA is expected to create a higher optical NR. Indeed, we find a greater optical NR in the driven 3LA than in a 2LA for a same asymmetry in the atom-photon coupling. We plot the NR in transmission coefficient $\Delta \mathcal{T}^{d0}$ through the driven 3LA in the absence of losses in Fig.~\ref{3LArec1} and compare it to the NR ($\Delta \mathcal{T}^{0}$) in the 2LA for a similar set of parameters. The maximum NR for the driven 3LA is 60$\%$ larger than that for the 2LA in Fig.~\ref{3LArec1}. The horizontal axis in Fig.~\ref{3LArec1} is a scaled intensity $I_{\rm in}/I^{\rm cr}_{\rm in}$ of an incident probe light where $I^{\rm cr}_{\rm in}$ is the critical intensity for the 2LA. We also observe that the NR in transmission occurs for a narrow  intensity range in the driven 3LA while it can occur for a wide intensity range in the 2LA. This is because the high optical nonlinearity in the driven 3LA is achieved in a narrow window controlled by $\delta \om_p$, $\Delta_c$ and $\Omega_c$. We also present a comparison in the NR between the driven 3LA and 2LA after inclusion of losses in Fig.~\ref{3LArec1} which shows that the NR ($\Delta \mathcal{T}^{d}$) is still higher in the driven 3LA. However, the NR is more sensitive to different losses in the driven 3LA than the 2LA.

Following the previous calculation for the 2LA, we next derive the contribution of coherent and incoherent scattering in the NR in transmission through the driven 3LA. In Fig.~\ref{3LArec2}, we show these contributions along with the total NR in transmission through the driven 3LA without any loss. Similar to the 2LA, we find here that while incoherent scattering has a larger contribution in NR at a higher intensity, the  maximum NR at a lower intensity is mainly due to coherent elastic scattering of the incident light.

\section{Conclusions and perspectives}
\label{conc}
Based on an exact microscopic analysis, we have shown several exciting features of nonreciprocal light transmission through two models of the nonlinear optical isolator. Specially, we have discussed some mechanisms to improve the NR in a nonlinear optical isolator. For a constant asymmetry in the coupling, the higher NR in a 3LA than a 2LA is directly related to the higher optical nonlinearity in the 3LA. Therefore, the NR can be treated as a measure of optical nonlinearity in these systems. Our theoretical analysis can be readily extended to more complex microscopic models of a nonlinear optical isolator such as in Refs.~\cite{RoyNatS2013, FratiniPRL2014, Shen14, Yu15, FratiniPRA2016}. It would also be interesting to check the effect of ultrastrong atom-light coupling on the NR for which we need to consider these systems beyond the Markov and rotating-wave approximations.  

\section*{Acknowledgments}
We thank C. M. Wilson, J. Hauschild, A. Vinu, J. Samuel, and R. Singh for discussion. The funding from the Department of Science and Technology, India via the Ramanujan Fellowship is gratefully acknowledged.

\bibliography{bibliography2}
\end{document}